\documentclass[12pt]{article}
\usepackage{geometry}
\usepackage{graphicx} 
\usepackage[textfont=bf, font = normalsize]{caption} 
\usepackage{floatrow} 
\usepackage[thinlines]{easytable}
\usepackage{amssymb}
\usepackage{amsfonts}
\usepackage{amsmath}
\usepackage[colorlinks]{hyperref}
\usepackage{subfig}
\usepackage{cite}
\usepackage{listings}
\usepackage{xcolor}
\floatstyle{plaintop} 
\restylefloat{table} 
\begin{document}

\title{\LARGE{ \bf Massive parallelization and performance enhancement of an immersed boundary method based unsteady flow solver}}
\author{Rahul Sundar\textsuperscript{1}, Dipanjan Majumdar\textsuperscript{1}, Chhote Lal Shah\textsuperscript{1},\\ and Sunetra Sarkar\textsuperscript{1}\\\\
\textsuperscript{1}\small{Department of Aerospace Engineering, IIT Madras, Chennai-600036, India}}
\date{}
\maketitle
\pagestyle{plain}

\noindent \textbf{ABSTRACT}

High-fidelity simulations of unsteady fluid flow are now possible with advancements in high performance computing hardware and software frameworks. Since computational fluid dynamics (CFD) computations are dominated by linear algebraic routines, they can be significantly accelerated through massive parallelization on graphics processing units (GPUs). Thus, GPU implementation of high-fidelity CFD solvers is essential in reducing the turnaround time for quicker design space exploration. In the present work, an immersed boundary method (IBM) based in-house flow solver has been ported to the GPU using OpenACC, a compiler directive-based heterogeneous parallel programming framework. Out of various GPU porting pathways available, OpenACC was chosen because of its minimum code intrusion, low development time, and striking similarity with OpenMP, a similar directive-based shared memory programming framework.
A detailed validation study and performance analysis of the parallel solver implementations on the CPU and GPU are presented. The GPU implementation shows a speedup up to the order $O(10)$ over the CPU parallel version and up to the order $O(10^2)$ over the serial code. The GPU implementation also scales well with increasing mesh size owing to the efficient utilization of the GPU processor cores.

\section{{\textbf{INTRODUCTION}}}
Unsteady flow past moving solid bodies is characterized by the flow field having a temporal dependence often manifesting through vortices surrounding and trailing the moving bodies\cite{lewin2003modelling,khalid2018bifurcations,lai1999jet,shelley2011flapping,young2007vortex}. These vortices stretch, rotate, merge, attach, and detach, such that each of these mechanisms have varying effects on the loads acting on the moving bodies immersed in a fluid\cite{bose2018investigating,majumdar2020capturing}. These mechanisms are also in return affected by displacement or deformation of the solid bodies\cite{shelley2011flapping}. 
To understand fundamental barriers to the optimal design of mechanical structures in flow, in-depth computational or experimental investigations are necessary. 

Although advancements in high-performance computing have made high-fidelity CFD simulations of unsteady flow possible, they are still computationally intensive\cite{alvarez214hierarchical, mader2020adflow, reguly2020productivity}. This is because the CFD algorithms are dominated by iterative linear algebra routines\cite{wood2020sparse} for the solution of Navier-Stokes (N-S) and allied governing equations. IBM \cite{peskin2002immersed,mittal2005immersed} forms one class of techniques in CFD that solves the flow governing equations on a grid non-conformal to the solid body immersed in the fluid. This avoids repetitive meshing otherwise necessitated by the traditional techniques such as arbitrary Lagrange eulerian (ALE) framework\cite{wick2013solving} which use a conformal grid. Thus, IBM is especially useful for simulations involving complex geometries\cite{kim2001immersed} and moving bodies\cite{mittal2005immersed}.

Recently, Majumdar \emph{et al.}\cite{majumdar2020capturing} developed a discrete forcing type IBM\cite{kim2001immersed} solver to simulate and capture dynamical transitions in unsteady flow past a sinusoidally plunging rigid elliptic foil in the low Reynolds number incompressible laminar flow regime. This solver was later parallelized by Shah \textit{et al.}\cite{shahperformance} using OpenMP, a compiler directive-based shared memory parallel programming framework, where a speed up of three times over the serial solver was reported. Although parallelization on CPUs using OpenMP allows for a reduction in turnaround time, these linear algebra routines can be accelerated further and significantly by offloading the computations onto GPUs\cite{xue2020heterogeneous}. This is because GPUs with their large number of processor cores and high throughput capacity promise massive performance enhancement of parallelizable linear algebra routines.

Typically, there exist three pathways to port a code to GPU: (i) performance tuned architecture specific third-party libraries - CuSPARSE and CuBLAS \cite{naumov2011incomplete} to name a few, (ii) compiler directive frameworks - OpenACC\cite{chandrasekaran2017openacc,farber2016parallel}, and (iii) architecture specific parallel programming language extensions such as CUDA\cite{hoshino2013cuda} and OpenCL\cite{sugawara2013comparison}. Of these pathways, compiler directive frameworks ensure minimum code intrusion and ease of parallelisation cutting down the development time\cite{li2016comparing} as compared to full-fledged parallel programming extensions like CUDA. 
 
The present work involves parallelization and performance enhancement of an in-house unsteady flow solver by offloading computationally intensive routines onto the GPU using the OpenACC framework. The paper outline is as follows: a discussion on the validation case setup and IBM solver is presented in section  \ref{sec2}. In section \ref{Sec:parstr}, baseline solvers chosen for validation, the code parallelization strategy, and the code development cycle are discussed. Performance analysis of the GPU ported solver in the context of overall speedup and input scaling is presented and compared with that of the baseline solvers in section \ref{Sec:results}. Finally in section \ref{Sec:conclsn}, conclusions and future work are discussed.

\section{\textbf{METHODOLOGY}}\label{sec2}
	\subsection{\textbf{Problem setup}}\label{Sec:valdnstd}
A sinusoidally plunging 2D rigid elliptic foil with a thickness-to-chord ratio of 0.12 immersed in a uniform free stream is considered in the present study as shown in Fig. \ref{Fig: Computational domain}(a). The non-dimensional plunging displacement ($\bar{y}(\bar{t})$) and velocity ($\dot{\bar{y}}(\bar{t})$) are given by,
\begin{align} \label{eq:kin1}
	\bar{y}(\bar{t}) &= \bar{h} \sin(k \bar{t}), \;\; \mbox{and}
	\\\dot{\bar{y}}(\bar{t}) &= k \bar{h} \cos(k \bar{t}). \label{eq:kin2}
\end{align}	
Here, $\displaystyle \bar{t} = \frac{tU_{\infty}}{c}$ is non-dimensional time,  $\displaystyle \bar{h} = \frac{h}{c}$ is non-dimensional plunging amplitude, where $t$ is dimensional time, $U_{\infty}$ is the free stream velocity, $h$ is the plunging amplitude and $c$ is chord length of the elliptic foil. Aligning with earlier literature \cite{khalid2018bifurcations,lewin2003modelling,majumdar2020capturing}, further discussions are based on the reduced frequency, $\displaystyle k = \frac{2\pi f_h c}{U_{\infty}}$ and the nondimensional plunging amplitude, $\bar{h}$, where, $f_h$ is the plunging frequency.

\subsection{\textbf{Immersed Boundary method}}
The present study assumes the flow to be laminar and governed by the incompressible N-S equations.
These equations are solved on a background Eulerian grid using discrete forcing IBM\cite{kim2001immersed}.
As a result of using an Eulerian grid with the solid body immersed in it, the  non-dimensional N-S equations take the form as follows,
\begin{equation}\label{eq:IBEq1}
	\frac{\partial \boldsymbol{u}}{\partial t} + \nabla.({\boldsymbol{u}}{\boldsymbol{u}})  = -{\nabla}{p} + \frac{1}{Re} {\nabla}^2{\boldsymbol{u}} + \boldsymbol{f},\;\;\;\;\mbox{and}
\end{equation}
\begin{equation}\label{eq:IBEq2}
	{\nabla}.{\boldsymbol{u}} - q = 0.
\end{equation}
Here, ${\boldsymbol{u}}$ represents the non-dimensional velocity vector with $u$ and $v$ being the horizontal and vertical components of $\boldsymbol{{u}}$. ${p}$ is non-dimensional pressure, and $Re$ is the Reynolds number given by $\displaystyle Re = \frac{U_{\infty}c}{\nu},$ where, $\nu$ is the kinematic viscosity.

While handling a non-conformal grid using IBM, the classification of fluid and solid mesh points is an added procedure at the beginning of every time marching step. 
The momentum forcing term $\boldsymbol{f}$ in Eq. (\ref{eq:IBEq1}) ensures the no-slip boundary condition on the solid boundary, and the source/sink term $q$ in Eq. (\ref{eq:IBEq2}) satisfies the continuity and minimise spurious force oscillations near the boundary~\cite{kim2001immersed}.
The flow equations (Eqs. (\ref{eq:IBEq1}) and (\ref{eq:IBEq2})) are solved using finite volume-based semi-implicit fractional step method\cite{choi1994effects} with the primitive variables being arranged in a staggered grid. Second-order spatial and temporal discretizations are achieved using Adams-Bashforth and Crank-Nicolson schemes respectively. 
Pressure and velocity correction equations are iteratively solved using a modified Gauss-Seidel successive over-relaxation method with red-black tagging as used by Shah \emph{et al.}\cite{shahperformance}. A detailed description of the IBM algorithm used in the present study is presented in Majumdar \emph{et al.}\cite{majumdar2020capturing}.

\subsection{\textbf{Computational domain and boundary conditions}}

A rectangular computational domain of size $[-7.5c,\; 24c] \times [-12.5c,\; 12.5c]$ is considered, with the elliptic foil placed at the origin initially. The grid size $\Delta x=\Delta y=0.004$ and time step $\Delta t=0.0001$ are chosen after performing grid and time convergence tests.
\begin{figure}[!htbp]
	\centering
	\captionsetup{width = \linewidth}	\includegraphics[width = 0.8
\linewidth, trim={0.0 0.0 0.2 0.2}]{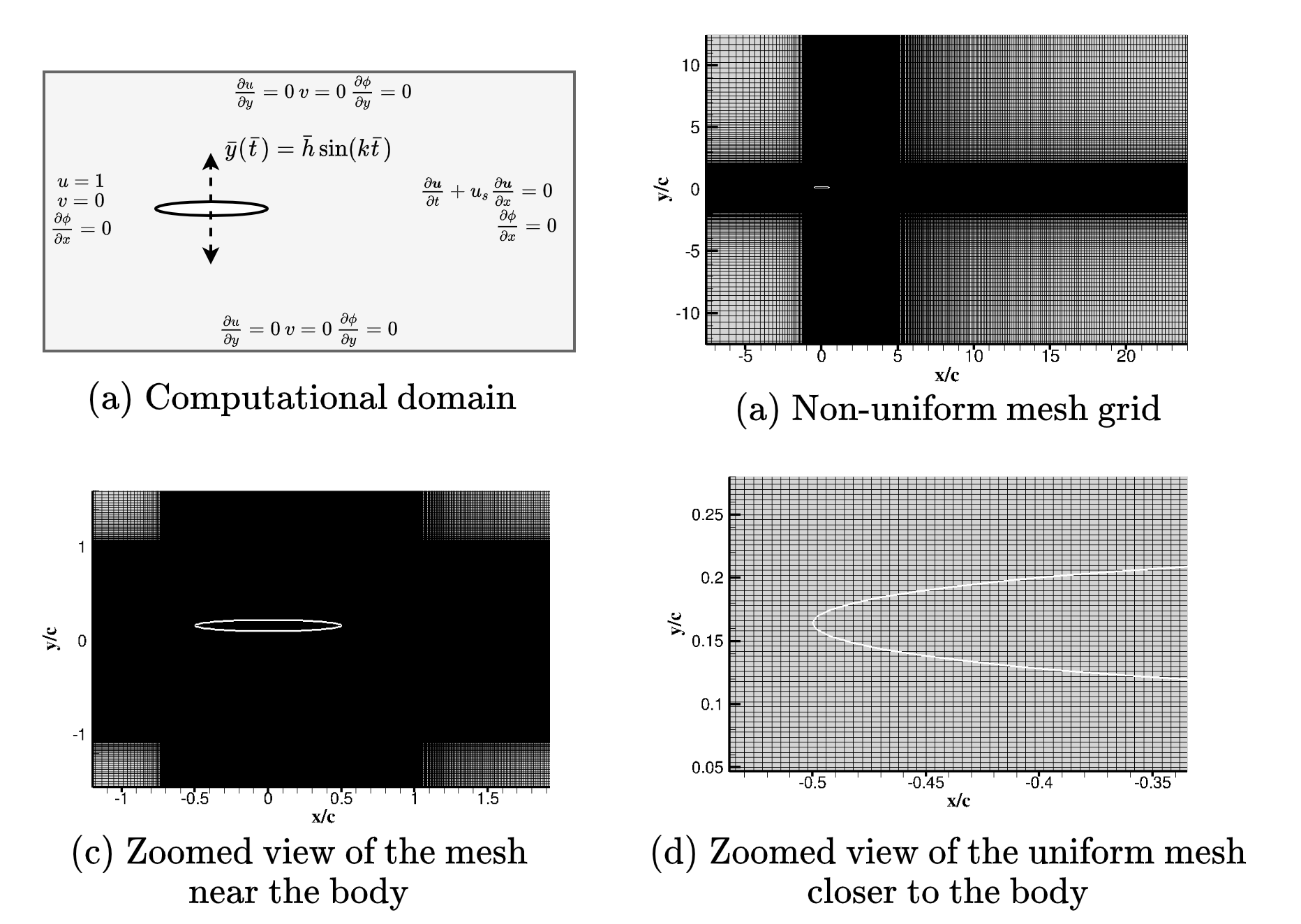}
	\caption{Computational domain}
	\label{Fig: Computational domain}
\end{figure}
The eulerian mesh is uniformly spaced in the region of the body movement with a minimum cell spacing and then gradually stretched following a geometric progression towards the outer boundaries as seen in Fig. \ref{Fig: Computational domain}(b) with the zoomed views presented in Figs. \ref{Fig: Computational domain}(c) and \ref{Fig: Computational domain}(d). 
 
\section{\textbf{Parallelisation strategy and implementation}}\label{Sec:parstr}
\subsection{\textbf{Earlier work and baseline solvers}}\label{Sec:baseline}
Two earlier versions of the in-house solver written in the C++ language are considered for validation and speedup analysis. These are, a serial implementation as in \cite{majumdar2020capturing}, and an OpenMP-based parallel implementation as in \cite{shahperformance}, henceforth referred to as \textbf{SOL0} and \textbf{SOL1}, respectively.
\textbf{SOL1} uses an improved Gauss-Seidel successive over-relaxation algorithm with red-black tagging of the mesh points to solve the pressure and velocity correction equations to avoid race conditions and data dependency. Owing to interspersed parallel and sequential regions, fork-join-based parallelism was adopted by Shah \emph{et al.}\cite{shahperformance}. As a result, a speedup of almost three times over \textbf{SOL0} was reported. The simulations using \textbf{SOL1} were executed on 16 CPU threads. However, the turnaround time was still high. Hence, there was a pertinent need to accelerate the solver by offloading already parallel or parallelizable computations to the GPU. In the current work, using the OpenACC framework, a GPU-compatible solver henceforth referred as \textbf{SOL2} is obtained. 

\subsection{\textbf{System configuration}}\label{Sec:sysconf}
Development and testing were first carried out on a local system with an Intel i7 10th Gen processor with 6 processor cores and 12 threads, 16Gb Memory, and a NVIDIA RTX 2060 Max Q GPU card with 1920 CUDA cores and 6GB memory. 
All the validation cases were however run on the AQUA super cluster hosted at the P.G Senapathy Computing Centre for Computing resource in IIT Madras. With a multithreaded 20-core Xeon Gold 6248 processor that has a clock speed of 2.6GHz, AQUA's GPU nodes consist of 2 NVIDIA Tesla V100 GPU cards each with 5120 CUDA cores and 32GB GPU memory.
\subsection{\textbf{Software stack}}\label{Sec:softw}
Throughout the present work, NVIDIA's high performance computing software development kernel (HPC-SDK) version - 20.7 has been used both on the local system and AQUA supercluster. Amongst a variety of libraries and tools in HPC-SDK 20.7, Nsight-Systems and Nsight-Compute were used here for code profiling and NVTX for code annotation. Portable Batch System (PBS) was used for job scheduling on the cluster.

\subsection{\textbf{GPU implementation}}\label{Sec:GPUim}
\begin{figure}[!htbp]
	\centering
	\captionsetup{width = \linewidth}
	\includegraphics[width=\linewidth]{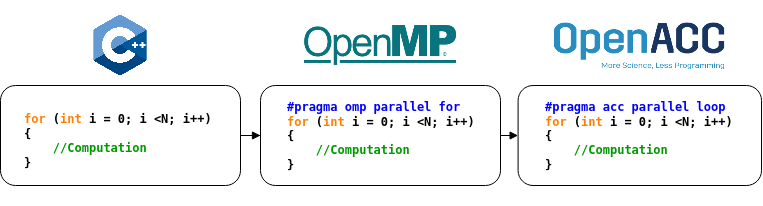}
	\caption{Schematics representing the minimal code changes when using OpenMP and OpenACC}
	\label{Fig: Code_Changes}
\end{figure}

The iterative code development cycle adopted in the present work using OpenACC to develop \textbf{SOL2} typically involves four steps: (i) analysis and profiling of the code for parallelizable hot spots, (ii) parallelization of the identified code hotspots, (iii) testing and validation of the modification, and (iv) further optimization of the parallelized loops. Since OpenMP and OpenACC are both compiler directive-based parallel programming frameworks, the GPU implementation involved very minimal source code intrusions as shown in the schematic in Fig. \ref{Fig: Code_Changes}.

Once the code sections are ported using appropriate OpenACC \textbf{pragmas}, the compiler automatically decides how to offload the code sections onto the GPU at the time of compilation. The implicit decisions made by the compiler can be inferred from the compiler trace. Further optimizations can also be made from the suggestions and feedback offered by the compiler trace. This is one reason why OpenACC has a quick learning curve and cuts down the code development time\cite{chandrasekaran2017openacc}. 
The code development cycle described earlier was followed for each code block in the parallelizable regions iteratively until a satisfactory performance was achieved. Details of the incremental GPU implementation are presented below. 

\subsubsection{\textbf{Analysis}}\label{Sec:ana}
{Using Nsight-systems, the serial CPU solver \textbf{SOL0} was initially profiled on the local system in which the functions: (i) flagging/classification of solid and fluid points, (ii) iterative solvers for pressure and velocity corrections were identified to be the most time-consuming regions followed by the code block involving (iii) body-force interpolation function as seen in Fig. \ref{Fig:Profile1} and Table \ref{tab:my_label}. The remaining regions of the code are executed serially and consume very little time as compared to the parallelizable sections.}

{As a preliminary analysis, the theoretical speedup $S_{theory}$ possible through parallelization of code regions over the serial execution is given by Amdahl's law, such that~\cite{amdahl1967validity}}
\begin{equation}
	S_{theory} = \frac{1}{(1-p) + p/n},
\end{equation}
where $n$ is the number of processor cores and $p = T_{par}/T_{wall}$ is the parallel portion of the code in terms of wall time taken for execution of the code respectively.  Here, $T_{ser}$ and $T_{par}$ are the time taken by serial and parallel sections of the code, respectively. Whereas, total wall time of the code, $T_{wall}$ is given by, $T_{wall} = T_{ser} + T_{par}. $ In an ideal scenario, with effective parallelization of the parallel portion, the total wall time $T_{wall}$ comes down closer to $T_{ser}$, with $T_{par}<< T_{ser}.$  In the present study, potentially parallelizable regions in \textbf{SOL0} comprise $99.8\%$ of the total wall time when executed serially, and hence, $p = T_{par}/T_{wall} = 0.998$.
{Thus, the maximum speedup possible in the limit of an extremely large number of processor cores (i.e $n\to\infty$, therefore $p/n\to 0$) would be $S_{theory} = 1/((1-0.998) + 0)  = 500$ times the sequential code.} However, the theoretical speedup is not always achievable owing to various factors such as a finite number of processor cores, hardware latency, data transfers between CPU and GPU, synchronization of threads, and lower processor clock speed of GPUs as compared to CPU. {Hence, the obtained speedup is much lesser. But, the effects of these factors can by minimized by profiling the code to identify the code hot spots needing optimization, and appropriate OpenACC pragmas can be added}. 

\begin{table}
    \caption{{Sequential CPU time for the functions identified to be parallelizable hot spots}}
    \centering
    \begin{tabular}{p{0.35\linewidth}  p{0.6\linewidth}}
        \textbf{Function} & \textbf{Computational time per time step (in seconds)} \\\hline\\
        P iterative solver & 3.835\\
        Fluid/solid flagging & 2.5515\\
        U-V iterative solver & 0.5518\\
        Body force interpolation & 0.0758\\\hline
    \end{tabular}
    \label{tab:my_label}
\end{table}

\subsubsection{\textbf{Parallelisation of loops}}\label{Sec:parloop}
\begin{figure}[!htbp]
	\centering
	\captionsetup{width = \linewidth}
	\includegraphics[width=0.8\linewidth]{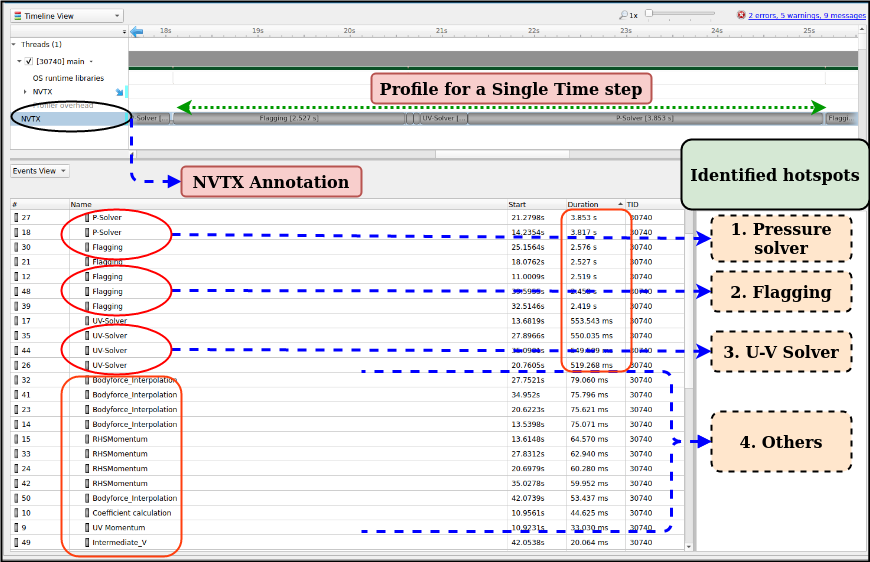}
	\caption{Nsight-systems profiler output for the NVTX annotated serial version of the IBM solver for a single time marching step.}
	\label{Fig:Profile1}
\end{figure}

{The parallelizable code blocks are often characterized by \textbf{for} loops. In the OpenMP implementation \textbf{SOL1}, the directive \textbf{parallel for} along with clauses \textbf{default(shared)} for data sharing of shared variables and \textbf{schedule(dynamic)} for balancing the workload distribution on the CPU threads were used for the hotspot regions earlier noted in Fig. \ref{Fig:Profile1} and in \cite{shahperformance}.} In the GPU implementation,
OpenACC directives such as \textbf{kernels} or \textbf{parallel for} can be added one by one before each potentially parallelizable code block /for loop/statement in the hotspots identified earlier (see Section \ref{Sec:ana}). The \textbf{kernels} directive is more flexible than the \textbf{parallel for} directive as it lets the compiler decide which regions within the scope of the construct are parallelizable and the compiler trace sheds more light on the specific optimizations and degrees of parallelism that can be brought in. To parallelize \textbf{for} loops with backward data dependencies, temporary variables can be allocated to swap and update the values. Although this would increase the memory requirement, it is offset by the improvement in speedup. Also, to avoid race conditions, selected variables in the scope of the code block that are offloaded onto GPU need to be appropriately privatized using the \textbf{private(variable list)} clause.

In the GPU implementation \textbf{SOL2}, classification/flagging of solid-fluid points function was first ported to GPU and then the pressure/velocity correction solver functions were ported using appropriate OpenACC directives and clauses. As mentioned in section \ref{Sec:ana}, body force interpolation procedure was also parallelized and offloaded onto the GPU in \textbf{SOL2}. 
 
\subsubsection{\textbf{Optimisation of loops}}\label{Sec:optloop}
Following the feedback from the compiler trace, appropriate directives/clauses were further added for optimization. Specifically, \textbf{routine} and  \textbf{seq} directives were added wherever there were sequential user-defined functions that needed to be executed on the GPU to avoid unnecessary data transfers between the CPU and the GPU. Also, care was taken to ensure that these offloaded sequential operations were not computationally intensive. Additionally, \textbf{collapse} and \textbf{reduction} clauses were added for vectorization and to avoid race conditions through simultaneous read and write operations respectively. 
Further optimizations such as explicit mapping of for loops to the GPU threads and other intricate data management constructs can also be implemented (for more details, refer to \cite{chandrasekaran2017openacc,xue2020heterogeneous}). However, in the present study, the code development cycle was stopped when a satisfactory speedup was obtained for \textbf{SOL2}, and the intricate optimizations were left for future work.

\subsubsection{\textbf{Data management}}\label{Sec:Datman}
\begin{figure}[!htbp]
	\centering
	\captionsetup{width = \linewidth}
	\includegraphics[width = 0.8\linewidth]{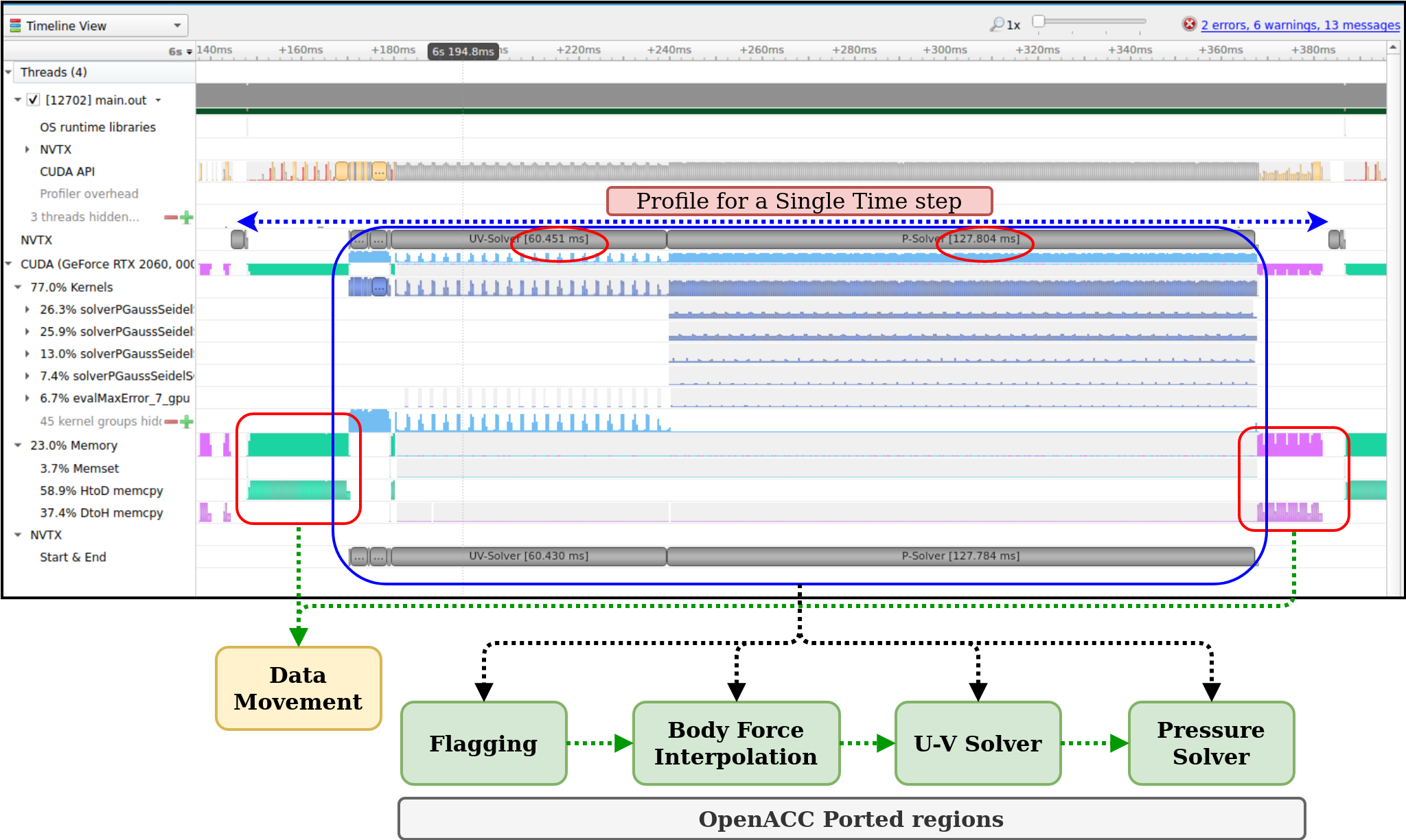}
	\caption{Final Nsight-systems profiler outputs for the NVTX annotated OpenACC version of the IBM solver for a single time marching step.}
	\label{Fig:Profile2}
\end{figure}

{Data management is crucial in OpenACC implementation as CPUs and GPUs have different memory architectures. Because of this, variables are first allocated on the CPU and then copied to GPU where computations are then carried out. This leads to multiple data transfers between CPU and GPU which gets limited by latency overheads and the memory bandwidth of GPU, thereby impeding the solver performance.} This can be avoided by minimizing data transfers using OpenACC's data management directive\cite{chandrasekaran2017openacc} appended with appropriate \textbf{copyin}, \textbf{copyout} or \textbf{copy} clauses. This drastically improves the performance because all computations can be carried out at once when the required data for all of them are hosted on the GPU at the time of execution.  In the present study, all the variables were copied to the GPU at the beginning of every time marching step of the solver. This is because file write operations that need to be carried out at the end of every few time steps take place only on the CPU. Hence, data transfers although significantly minimized, were necessary at the beginning and end of every time step for \textbf{SOL2} as seen in Fig. \ref{Fig:Profile2}.

\section{\textbf{Results and discussion}}\label{Sec:results}
\subsection{\textbf{Validation studies}}\label{Sec:Valres}	
\begin{figure}
	\centering
	\captionsetup{width = \linewidth}
	\subfloat[]{\includegraphics[width = 0.7\linewidth]{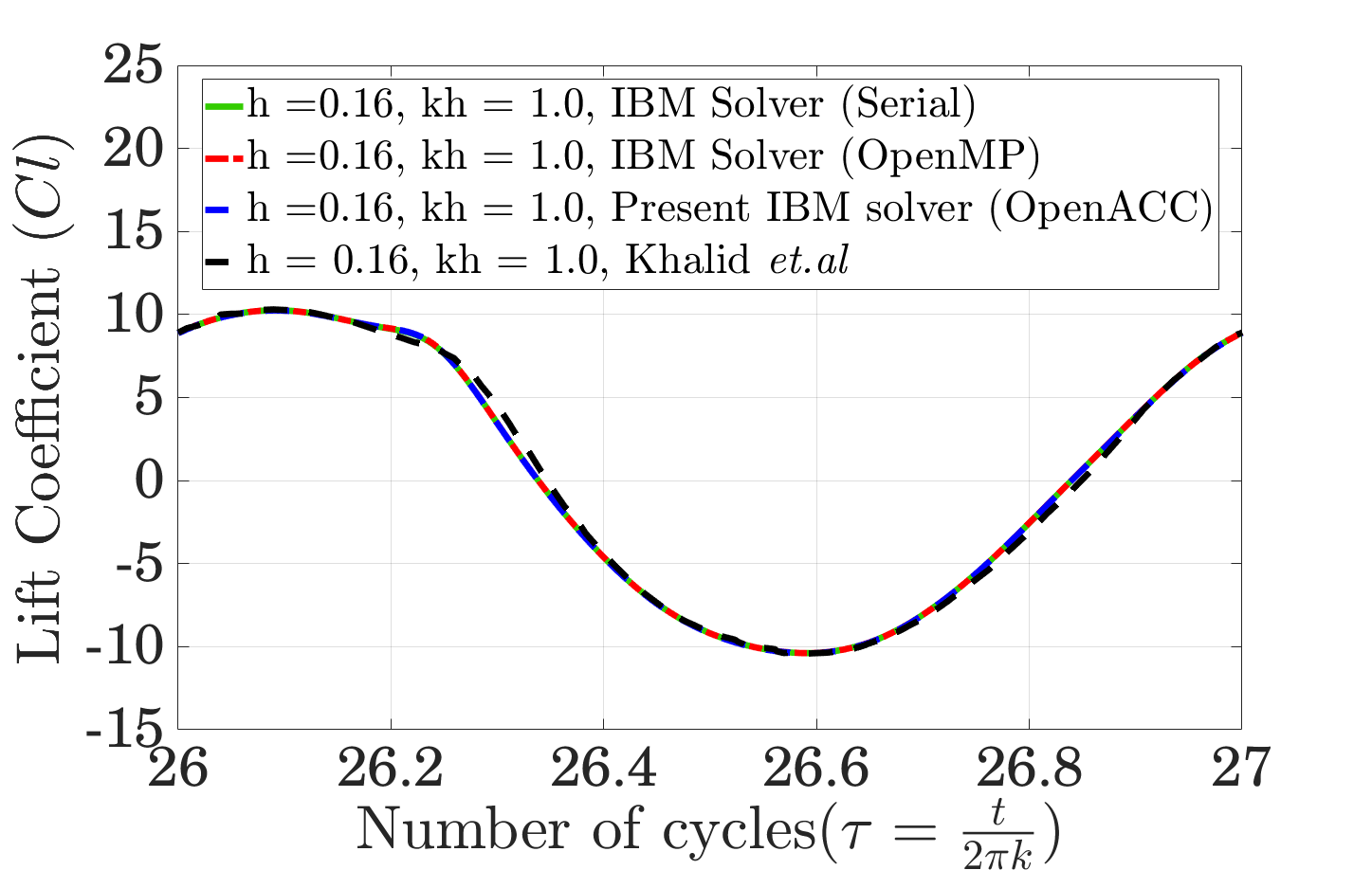}}\\
	\subfloat[]{\includegraphics[width = 0.7\linewidth]{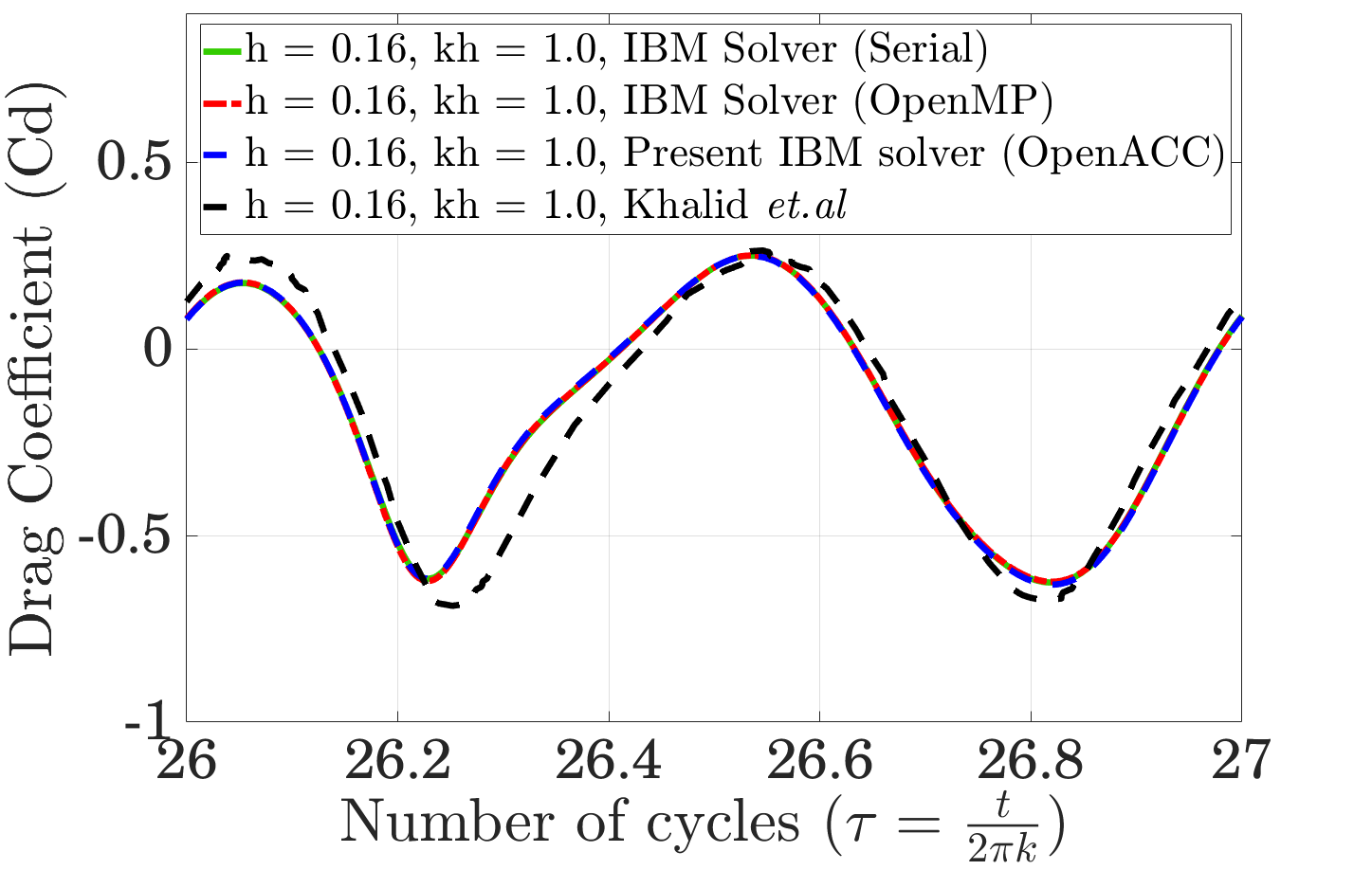}}
	\caption{Plots comparing the (a) lift and (b) drag coefficient time histories for serial, OpenMP, OpenACC implementations with the results of Khalid \emph{et al.}\cite{khalid2018bifurcations} for a sinusoidally plunging rigid elliptic foil.}
	\label{Fig:Aerodynamic_coeffs}
\end{figure}

A Reynolds number, $Re = 500$, non-dimensional plunging velocity $k\bar{h} = 1.0$ with $k= 2\pi$ and $\bar{h} = 0.16$ are chosen for validation. The aerodynamic force coefficient time histories obtained from the simulations were compared with that of the baseline solvers \textbf{SOL0}, \textbf{SOL1} and the work of Khalid \emph{et al.}\cite{khalid2018bifurcations} in Fig. \ref{Fig:Aerodynamic_coeffs}. The results are in good agreement with each other for \textbf{SOL0}, \textbf{SOL1} and \textbf{SOL2} solvers with almost no deviations, and also corroborate well with the results of Khalid \emph{et al.}\cite{khalid2018bifurcations}. The small discrepancies observed in the aerodynamic coefficients between the present numerical studies and the literature\cite{khalid2018bifurcations}, especially in drag coefficient (see Fig. \ref{Fig:Aerodynamic_coeffs}(b)), are attributed to underlying differences in the numerical method used in \cite{khalid2018bifurcations} and \cite{majumdar2020capturing}.           
\subsection{\textbf{Performance analysis}}\label{Sec:Perf}
\subsubsection{\textbf{Speedup characterisation}} \label{Sec:sped}
\begin{figure}[!ht]
	\centering
	\captionsetup{width = \linewidth}
	\subfloat[]{\includegraphics[width = 0.7\linewidth]{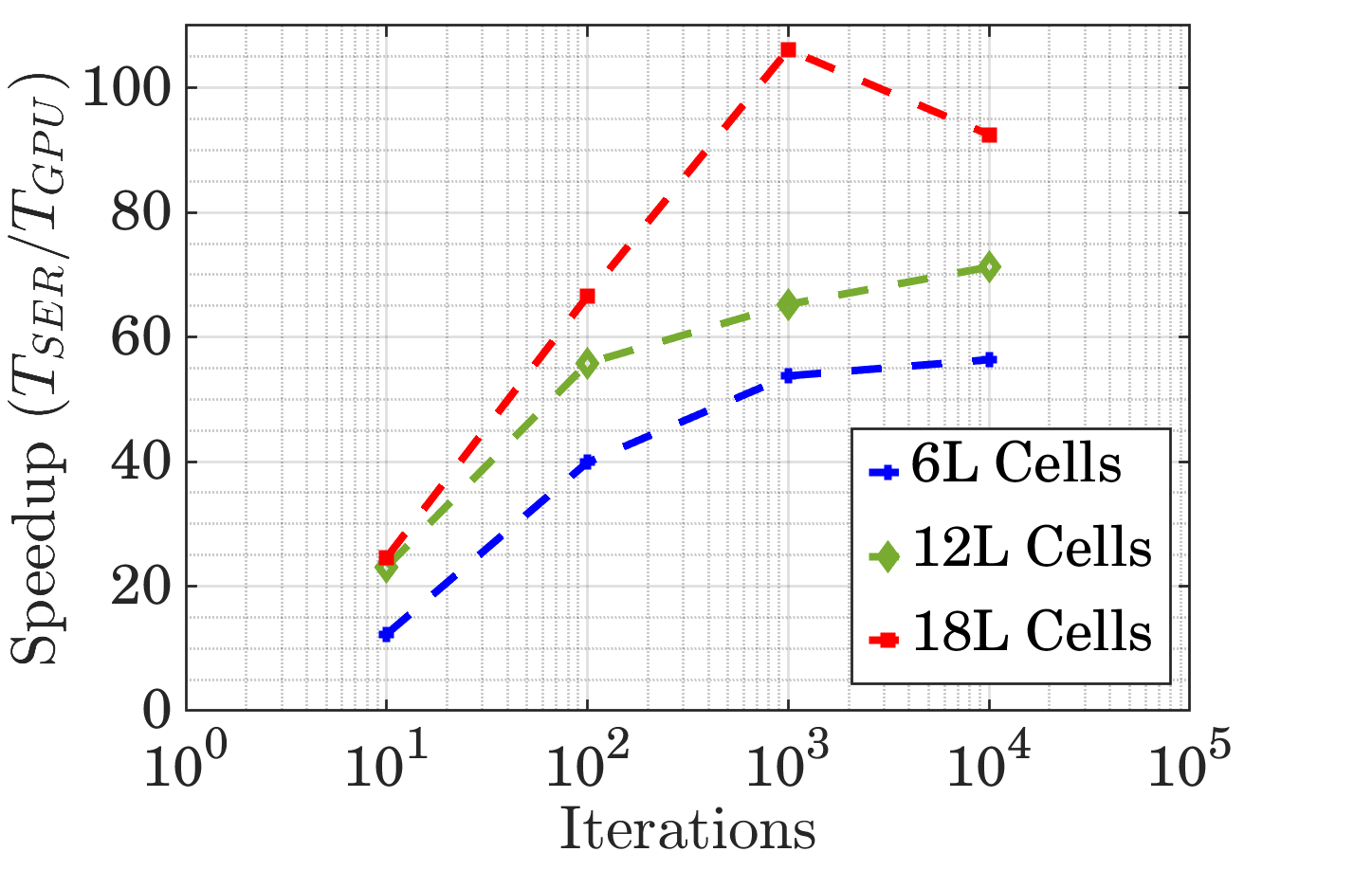}}\\
	\subfloat[]{\includegraphics[width = 0.7\linewidth]{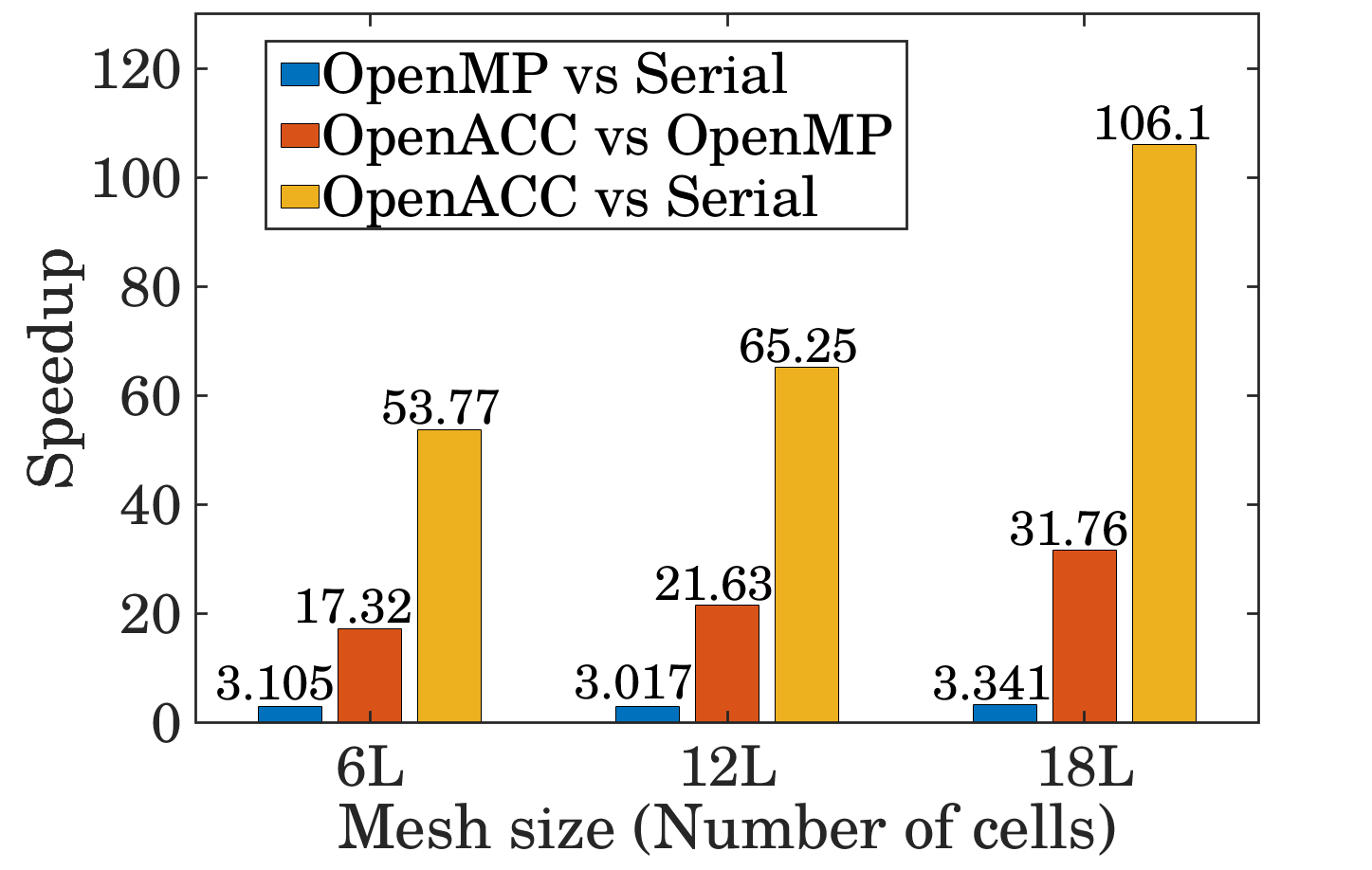}}
	\caption{(a) Evolution of speedup  as a function of timestep for the GPU ported IBM solver at various mesh levels, and (b) bar graph depicting the relative speedup of OpenACC version obtained over serial and openMP versions at different mesh levels and for the first 1000 time steps}
	\label{Fig: performance}
\end{figure}

Cases for each solver setting are run thrice to obtain the average wall times for \textbf{SOL0}, \textbf{SOL1}, and \textbf{SOL2}. To characterize the speedup of the parallel solvers \textbf{SOL1} and \textbf{SOL2}, the wall times of \textbf{SOL0} are considered as references. The speedup $S$ can be calculated using the wall times as follows
\begin{equation}
	S_{SOLi} = T_{SOL0}/T_{SOLi} \;\; \mbox{for i = 1,2} 
\end{equation}
The relative speedup obtained by \textbf{SOL2} over \textbf{SOL1} is calculated as follows
\begin{equation}
	S_{relative} = S_{SOL2}/S_{SOL1}
\end{equation}
Additionally, the evolution of average speedup of \textbf{SOL2} over \textbf{SOL0} with the increasing number of time steps is considered for the first $10, 100, 1000$ and $10000$ time steps respectively (see Fig. \ref{Fig: performance}(a)). On considering the overall speedup for the first 1000 time steps alone, it is observed that a significant reduction, in turn, around times (see Table \ref{Tab:WallTime}) is achieved that results in almost a speedup of the order $O(10^2)$ and $O(10)$ over \textbf{SOL0} and \textbf{SOL1}, respectively.

\begin{table}[!ht]
    \renewcommand{\arraystretch}{2}
	\centering
	\captionsetup{width = \linewidth}
	\begin{tabular}{llll}
		\hline
		Mesh Levels  & M1 (6L) & M2 (12L) & M3 (18L) \\
		\textbf{Solver version}  & &  & \\
		\hline
		\textbf{SOL0} & 13140s & 24663s & 39994s \\
		\textbf{SOL1} & 4232s & 8175s & 11953s  \\
		\textbf{SOL2} & \textbf{244.4s} & \textbf{378s} & \textbf{368.3s} \\
		\hline
	\end{tabular}
	\caption{Average wall time for serial CPU, OpenMP and OpenACC GPU solvers executed for the first 1000 time steps}
	\label{Tab:WallTime}
\end{table}
\subsubsection{\textbf{Input scaling performance}}\label{Sec:Inpsc}
Along with speedup, given the massive number of GPU cores, it is important that scaling of speedup concerning increasing mesh levels of the computation domain is studied. To this effect, three levels of mesh were considered, M1, M2, and M3 with 6, 12, and 18 Lakh grid cells respectively. 
In Fig. \ref{Fig: performance}(a), as the number of time steps are increased, a flattening trend is observed in the case of meshes M1 and M2 but in the case of M3, the speedup scales linearly with time up to 1000 time steps and then dips suddenly. The significant difference in speedup pattern for the mesh M3 compared to M1 and M2 is attributed to the uncertainty in job allocation on the GPU nodes of AQUA cluster. The jobs pertaining to M1 and M2 were often run on one GPU node while those of M3 ran on another GPU nodes with multiple concurrent jobs running simultaneously. 

The overall and relative speedup obtained for \textbf{SOL2} over \textbf{SOL1} and \textbf{SOL0} for the first 1000 time steps are presented for the three mesh levels in \ref{Fig: performance}(b). 
The OpenACC-based solver scales almost linearly with increasing mesh size compared to the sequential and OpenMP versions. This indicates the effective utilization of GPU cores with increasing mesh size. This is in contrast to the OpenMP version of the solver where speedup over serial code is almost constant across mesh levels.

\section{\textbf{Conclusion}}\label{Sec:conclsn}
To reduce the turnaround time of an IBM-based unsteady flow solver, OpenACC was chosen as the GPU porting pathway owing to its similarity with OpenMP, minimal code intrusion, and development time. Computationally intensive parallel routines were offloaded onto the GPU with minimal data transfers using appropriate directives and clauses provided in the OpenACC framework by adopting an incremental code development cycle.
Finally, significant speedups up to the order $O(10)$ and $O(10^2)$ over the OpenMP and serial solver versions were obtained, respectively. Improved performance of the GPU ported solver is a result of (i) parallelized code regions that were otherwise executed sequentially in the openMP version, (ii) optimized code blocks / for loops/functions with specific directives and clauses ensuring no data dependency, data conflicts, and race condition, and (iii) minimized data transfers between CPU and GPU.  With increasing mesh size, the speedup of OpenACC version scaled almost linearly owing to the effective utilization of the GPU cores as compared to the constant speedup of the OpenMP version of the code. 
Extension of the parallelization strategy followed in the present work to other in-house IBM-based non-linear fluid-structure interaction solvers being developed with a similar performance analysis and validation is left for future work. 

\vspace{0.5cm}
\noindent
\textbf{ACKNOWLEDGEMENTS}\\
\noindent We thank the HPCE team of IIT Madras for providing us with computational resources of AQUA super cluster. We also thank CDAC-Pune and NVIDIA for the guidance provided during the CDAC-GPU hackathon 2020, organized under the aegis of National Super Computing Mission, Govt. of India.

\clearpage
\bibliographystyle{unsrt}
\bibliography{bibliography.bib}

\end{document}